
\def\gsim{\ \rlap{\raise 2pt \hbox{$>$}}{\lower 2pt \hbox{$\sim$}}\ }
\def\lsim{\ \rlap{\raise 2pt \hbox{$<$}}{\lower 2pt \hbox{$\sim$}}\ }

\rightline{hep-ph/9507290, WIS-95/28/Jul-PH}
\vskip 2cm
{\vbox{\centerline{\bf Exploring New Physics with CP Violation in
Neutral $D$ and $B$ Decays}}}
\bigskip
\bigskip
\centerline{Yosef Nir}
\smallskip
\centerline{\it Department of Particle Physics}
\centerline{\it Weizmann Institute of Science, Rehovot 76100, Israel}
\bigskip
\bigskip
\baselineskip 18pt

\noindent
If New Physics contributes significantly to neutral meson mixing,
then it is quite likely that it does so in a CP violating manner.
In $D^0-\bar D^0$ mixing measured through $D^0\rightarrow
K^+\pi^-$, CP violation induces a term $\propto te^{-\Gamma t}$ with
important implications for experiments.
For $B_s-\bar B_s$ mixing, a non-vanishing CP asymmetry
(above a few percent) $a_{CP}(B_s \rightarrow D_s^+ D_s^-)$
is a clear signal of New Physics. Interestingly,
this would test precisely the same Standard Model ingredients as
the question of whether $\alpha+\beta+\gamma=\pi$.

\bigskip
\bigskip
\centerline{Talk presented at the}
\centerline{6th International Symposium on Heavy Flavour Physics}
\centerline{Pisa, Italy\ \ \ \ \ June 6--10, 1995}
\endpage

\REF\bsn{G. Blaylock, A. Seiden and Y. Nir, hep-ph/9504306,
Phys. Lett. B, in press.}
\REF\nisi{Y. Nir and D. Silverman, Nucl. Phys. B345 (1990) 301.}
\REF\nirev{Y. Nir, in {\it Proceedings of the Workshop on
  B Physics at Hadron Accelerators}, eds. P. McBride and C.S. Mishra,
  SSCL-SR-1225, Fermilab-CONF-93/267 (1993), p.185;\hfill\break
Y. Nir and H.R. Quinn, Ann. Rev. Nucl. Part. Sci. 42 (1992) 211.}
\REF\expt{J.C. Anjos {\it et al.}, Phys. Rev. Lett. 60 (1988) 1239.}
\REF\nirrev{Y. Nir, in {\it Proc. of the 20th Annual Slac Summer
Institute on Particle Physics:  The Third Family and the Physics of
Flavor}, Stanford, CA (1992), p. 81.}
\REF\Bigi{I.I. Bigi, in {\it Proc. of the 23rd International
Conference on High Energy Physics}, ed. S.C. Loken, Singapore,
World Scientific (1986), p.857; {\it Charm Physics}. ed. M. Ye
and T. Huang, N.Y., Gordon and Breach (1987), p.339.}
\REF\Liu{T. Liu, preprint HUTP-94/E021, hep-ph/9408330.}
\REF\guy{G. Blaylock, private communication.}
\REF\BrPa{T. Browder and S. Pakvasa, in preparation.}
\REF\wolf{L. Wolfenstein, preprint CMU-HEP95-05, hep-ph/9505285.}
\REF\QSA{Y. Nir and N. Seiberg, Phys. Lett. B309 (1993) 337;\hfill\break
M. Leurer, Y. Nir and N. Seiberg, Nucl. Phys. B420 (1994) 468.}
\REF\fourth{K.S. Babu, X.-G. He, X.-Q. Li and S. Pakvasa,
 Phys. Lett. B205 (1988) 540.}
\REF\Usinglet{G.C. Branco, P.A. Parada and M.N. Rebelo,
 preprint UWThPh-1994-51, hep-ph/9501347.}
\REF\LQ{M. Leurer, Phys. Rev. Lett. 71 (1993) 1324;
 Phys. Rev. D48 (1994) 333;\hfill\break
S. Davidson, D. Bailey and B.A. Campbell, Z. Phys. C61 (1994) 613.}
\REF\NFC{L.F. Abbott, P. Sikivie and M.B. Wise,
 Phys. Rev. D21 (1980) 1393;\hfill\break
V. Barger, J.L. Hewett and R.J.N. Phillips,
 Phys. Rev. D41 (1990) 3421;\hfill\break
Y. Grossman, Nucl. Phys. B426 (1994) 355.}
\REF\noNFC{S. Pakvasa and H. Sugawara, Phys. Lett. B73 (1978) 61;
\hfill\break T.P. Cheng and M. Sher, Phys. Rev. D35 (1987) 3484;
\hfill\break L. Hall and S. Weinberg, Phys. Rev. D48 (1993) R979.}
\REF\niqu{Y. Nir and H.R. Quinn, Phys. Rev. D42 (1990) 1473.}

\chapter{Introduction}
If New Physics contributes significantly to neutral meson mixing,
then it is quite likely that it does so in a CP violating manner.
This could have important consequences:
\item{a.} In $D^0-\bar D^0$ mixing measured through $D^0\rightarrow
K^+\pi^-$, a relative phase between the direct decay amplitude and
the mixing amplitude induces a term $\propto te^{-\Gamma t}$ with
important implications for experiments.
\item{b.} In $B^0-\bar B^0$ mixing, the theoretical calculation of
the mixing suffers from large hadronic uncertainties that makes it
difficult to uncover contributions from New Physics. In contrast,
in CP asymmetries in neutral $B$ decays into final CP eigenstates,
{\it e.g.} $a_{CP}(B\rightarrow\psi K_S)$, the hadronic uncertainties
are small and new CP violating contributions to mixing may be clearly
signalled.
\item{c.} In $B_s-\bar B_s$ mixing, the Standard Model CP violating
phase in the mixing amplitude is,
to a good approximation, equal to that of the $b\rightarrow c\bar cs$
decay amplitude. Consequently, a non-vanishing CP asymmetry
(above a few percent) $a_{CP}(B_s \rightarrow D_s^+ D_s^-)$
is a clear signal of New Physics. Interestingly,
this would test precisely the same Standard Model ingredients as
the question of whether $\alpha+\beta+\gamma=\pi$.

In section 2 we study the role of CP violation in $D-\bar D$ mixing.
The content of this section follows ref. [\bsn], but benefits from
the very useful discussions with several colleagues,
particularly Sandip Pakvasa and Guy Blaylock.
In section 3 we prove the relation between $a_{CP}(B_s\rightarrow
D_s^+ D_s^-)$ and the relation $\alpha+\beta+\gamma=\pi$.
The content of this section is based on ref. [\nisi], but the
presentation is different. The investigation of CP asymmetries in
$B^0$ decays as a probe of New Physics has been recently reviewed
in ref. [\nirev] and is not repeated here.

\chapter{CP Violation in Neutral $D$ decays}
The best bounds on $D-\bar D$ mixing come from measurements
of $D^0\rightarrow K^+\pi^-$ [\expt]. However, these bounds are still
orders of magnitude above the Standard Model prediction for the
mixing. If the value of $\Delta m_D$ is anywhere close to
present bounds, it should be dominated by New Physics.
Then, new CP violating phases may play an important role in
$D-\bar D$ mixing. In this section, we investigate the
consequences of CP violation from New Physics in neutral $D$ mixing.

There are three types of CP violation in meson decays [\nirrev]:
in decay, in mixing and in the interference of mixing and decay.
We first argue that only CP violation in the interference of
mixing and decay is likely to be relevant in the experimental search
for $D-\bar D$ mixing through $D^0\rightarrow K^+\pi^-$.

(i) {\it CP Violation in decay}: The decay $D^0\rightarrow K^+\pi^-$
proceeds via the quark process $c\rightarrow d\bar su$.
Within the Standard Model, this is completely dominated by
doubly Cabibbo suppressed (DCS) tree amplitudes. There is
no reasonable type of New Physics that could contribute to
charm decays comparably to the $W$-mediated diagram. Consequently,
$D^0\rightarrow K^+\pi^-$ is dominated by the single weak phase
$\arg(V_{us}V_{cd}^*)$. Similarly, the Cabibbo-allowed
mode, $D^0\rightarrow K^-\pi^+$ is dominated by a single weak phase,
$\arg(V_{ud}V_{cs}^*)$. It is very safe to assume that there is
no CP violation in decay for these modes.

(ii) {\it CP Violation in mixing}: For the neutral $D$ mass
eigenstates to differ from the CP eigenstates, one has to have
Im$(\Gamma_{12}/M_{12})\neq0$. If $\Delta m_D$ is anywhere close
to present bounds, then it is clearly dominated by New Physics,
$M_{12}\gg M_{12}^{\rm SM}$. On the other hand, there is no reasonable
type of New Physics that could enhance $\Gamma_{12}$ by orders of
magnitude, so  that very likely $\Gamma_{12}\sim\Gamma_{12}^{\rm SM}$.
Therefore, if $\Delta m_D$ is close to the present bounds,
it is very safe to assume that there is no CP violation in mixing.
(This assumption may have to be dropped if experiments reach
the sensitivity close to the Standard Model estimate.)

(iii) {\it CP Violation in the interference of mixing and decay}:
Within the Standard Model, both the mixing amplitude for neutral
$D$ mesons and the decay amplitude for $D\rightarrow K\pi$ occur
through processes that involve, to a very good approximation,
quarks of the first two generations only. Therefore, the relative
weak phase between the mixing and decay amplitudes is extremely small.
However, most if not all extensions of the Standard Model that allow
$\Delta m_D$ close to the limit involve new CP violating phases.
In these models, the relative phase between the mixing amplitude
and the decay amplitude is usually unconstrained and would naturally
be expected to be of ${\cal O}(1)$. (Examples are given below.)
CP violation of this type could then be a large effect.

We now investigate the implications of the fact that CP violation
in the interference of mixing and decay could be an effect of
${\cal O}(1)$ and, moreover, that other types of CP violation
are negligibly small. To do that, we first introduce some
formalism and notations (see also discussions in [\Bigi,\Liu]).

We define $p$ and $q$ as the strong interaction eigenstate components
in the mass eigenstates $\ket{D_{1,2}}$:
$$\ket{D_{1,2}}=p\ket{D^0}\pm q\ket{\bar D^0}.\eqn\defpq$$
Denoting the masses and widths of $D_{1,2}$ by $M_{1,2}$ and
$\Gamma_{1,2}$, we define their sums and differences:
$$\eqalign{
M\equiv{1\over2}(M_1+M_2),&\ \ \ \Delta M\equiv M_2-M_1,\cr
\Gamma\equiv{1\over2}(\Gamma_1+\Gamma_2),&\ \ \
\Delta\Gamma\equiv\Gamma_2-\Gamma_1.\cr}\eqn\sumdiff$$
We define the four decay amplitudes
$$\eqalign{
A\equiv\VEV{K^+\pi^-|H|D^0},\ \ \
B\equiv\VEV{K^+\pi^-|H|\bar D^0},\cr
\bar A\equiv\VEV{K^-\pi^+|H|\bar D^0},\ \ \
\bar B\equiv\VEV{K^-\pi^+|H|D^0}.\cr}\eqn\defAB$$
Finally, we define the phase convention independent quantities
$$\lambda={p\over q}{A\over B},\ \ \
\bar\lambda={q\over p}{\bar A\over\bar B}.\eqn\deflam$$

Our discussion above of CP violation has the following implications:
\item{(i)} As CP violation in decay is negligible,
$${|A|\over|\bar A|}={|B|\over|\bar B|}=1.\eqn\noCPd$$
\item{(ii)} As CP violation in mixing is negligible,
$$\left|{p\over q}\right|=1.\eqn\noCPm$$

Eqs. \noCPd\ and \noCPm\ together imply also $|\lambda|=|\bar\lambda|$.
Furthermore, the following approximations can be safely made:
\item{(iii)} We will assume -- as confirmed experimentally -- that
$\Delta M\ll\Gamma$, $\Delta\Gamma\ll\Gamma$ and $|\lambda|\ll1$.
\item{(iv)} We will also take here $\Delta\Gamma\ll\Delta M$,
which is very likely if $\Delta M$ is close to the bound.

The consequence of $(i)-(iv)$ is the following
form for the (time dependent) ratio between the DCS and
Cabibbo-allowed decay rates ($D^0(t)$ [$\bar D^0(t)$] is the
time-evolved initially pure $D^0$ [$\bar D^0$] state):
$$\eqalign{
{\Gamma[D^0(t)\rightarrow K^+\pi^-]\over
\Gamma[D^0(t)\rightarrow K^-\pi^+]}=&\
|\lambda|^2+{\Delta M^2\over4}t^2+{\rm Im}(\lambda)\ t,\cr
{\Gamma[\bar D^0(t)\rightarrow K^-\pi^+]\over
\Gamma[\bar D^0(t)\rightarrow K^+\pi^-]}=&\
|\lambda|^2+{\Delta M^2\over4}t^2+{\rm Im}(\bar\lambda)\ t.\cr}
\eqn\master$$
This form is valid for time $t$ not much larger than ${1\over\Gamma}$.
The time independent term is the DCS decay contribution;
the term quadratic in time is the pure mixing contribution;
and the term linear in time results from the interference
between the DCS decay and the mixing amplitudes.
Note that both the const($t$) and the $t^2$ terms are equal
in the $D^0$ and $\bar D^0$ decays. However, if CP violation
in the interference of mixing and decay is significant,
${\rm Im}(\lambda)\neq{\rm Im}(\bar\lambda)$ is possible,
and the linear term may be different for $D^0$ and $\bar D^0$.

The experimental strategy should then be as follows [\guy,\BrPa]:
(a) Measure $D^0$ and $\bar D^0$ decays separately.
(b) Fit each of the ratios to constant plus linear plus quadratic
time dependence.
(c) Combine the results for $|\lambda|^2$ and $\Delta M^2$.
(d) Compare Im($\lambda$) to Im($\bar\lambda$).

The comparison of the linear term should be very informative
about the interplay between strong and weak phases in these decays.
There are four possible results:
\item{1.} ${\rm Im}(\lambda)={\rm Im}(\bar\lambda)=0$:
Both strong phases and weak phases play no role in these processes.
\item{2.} ${\rm Im}(\lambda)={\rm Im}(\bar\lambda)\neq0$:
Weak phases play no role in these processes. There is a different
strong phase shift in $D^0\rightarrow K^+\pi^-$ and
$D^0\rightarrow K^-\pi^+$.
\item{3.} ${\rm Im}(\lambda)=-{\rm Im}(\bar\lambda)$:
Strong phases play no role in these processes. CP violating phases
affect the mixing amplitude.
\item{4.} $|{\rm Im}(\lambda)|\neq|{\rm Im}(\bar\lambda)|$:
Both strong phases and weak phases play a role in these processes.

In all these cases, the magnitude of the strong and the weak phases
can be determined from the values of $|\lambda|$, Im($\lambda)$
and Im($\bar\lambda$).

Finding either quadratic or linear time dependence would be
a signal for mixing in the neutral $D$ system. However,
a non-vanishing linear term does not by itself signal CP violation
in mixing, only if it is different in $D^0$ and $\bar D^0$.
The linear term could be a problem for experiments: if the phase
is such that the interference is destructive, it could partially
cancel the quadratic term in the relevant range of time, thus
weakening the experimental sensitivity to mixing [\bsn]. On the other
hand, if the mixing amplitude is smaller than the DCS one,
the interference term may signal mixing even if the pure mixing
contribution is below the experimental sensitivity [\Liu,\wolf].

Before concluding, we briefly survey some types of New Physics
that allow large $D-\bar D$ mixing and the source of CP violation
in each of them that allows large CP violation in the interference of
neutral $D$ mixing and $D\rightarrow K\pi$ decay.

{\bf Supersymmetry with quark--squark--alignment} [\QSA] is a unique
class of models in that it not only allows but actually requires
$\Delta m_D$ close to the bound. Large $\Delta m_D$ comes from
box diagrams with intermediate gluinos and up and charm squarks.
The mixing matrix for gluino--quark--squark couplings has
new CP violating phases (not related to the CKM matrix)
so that the phase of the mixing amplitude is arbitrary.

{\bf Fourth quark generation} [\fourth] contributes to $\Delta m_D$
through box diagrams with intermediate $b^\prime$ quarks.
The $4\times4$ charged current mixing matrix has three
CP violating phases
so that the phase of the mixing amplitude is arbitrary.

{\bf Left-handed $SU(2)$-singlet up quarks} [\Usinglet] allow
the $Z$-boson to couple non-diagonally to the up sector
(and, similarly, right-handed $SU(2)$ doublet up quarks).
Large $\Delta m_D$ may come from $Z$-mediated tree diagrams.
The neutral-current mixing matrix has new CP violating phases
(related to new phases in the charged current mixing matrix)
so that the phase of the mixing amplitude is arbitrary.

{\bf Light scalar leptoquarks} [\LQ] contribute to $\Delta m_D$
through box diagrams with intermediate leptons. Scalar
leptoquark couplings carry arbitrary new phases
so that the phase of the mixing amplitude is arbitrary.

{\bf Multi-scalar models with natural flavor conservation}
[\NFC] introduce a charged Higgs that may contribute to $\Delta m_D$
through box diagrams similar to the SM but with one or two of
the $W$ propagators replaced by the charged Higgs. If the
diagram with intermediate $b$ quark is large enough, its contribution
$\propto V_{ub}^*V_{cb}$ allows the CKM phase to affect
$D-\bar D$ mixing.

{\bf Multi-scalar models without natural flavor conservation}
[\noNFC] allow neutral scalars to couple non-diagonally to quarks.
A large contribution to $\Delta m_D$ is possible from
scalar mediated tree diagram. The couplings of the scalar may
depend on arbitrary new phases, though such phases may give
a too large contribution to $\epsilon_K$.

In summary, various extensions of the Standard Model allow
large, CP violating, contributions to $D-\bar D$ mixing.
This will induce an interference term between the DCS contribution
and the mixing contribution to $D^0\rightarrow K^+\pi^-$.
While such a term may be the consequence of strong phase shifts,
a CP violating contribution will be unambiguously signalled if
it is different in $D^0\rightarrow K^+\pi^-$ and
$\bar D^0\rightarrow K^-\pi^+$.

\chapter{What Does $\alpha+\beta+\gamma=\pi$ Test?}

It is often stated that whether the angles $\alpha$, $\beta$ and
$\gamma$ measured by the CP asymmetries in {\it e.g.}
$B\rightarrow\psi K_S$, $B\rightarrow\pi\pi$, and
$B_s\rightarrow\rho K_S$, respectively, fulfill
$$\alpha+\beta+\gamma=\pi\eqn\sumpi$$
will be a stringent test of the Standard Model.
We here wish to show that [\nisi]
\item{a.} If \sumpi\ is violated, it will be a clean indication
that $B_s$ mixing is {\it not} dominated by the Standard Model
box diagrams, and
\item{b.} Precisely the same information will be provided by the much
simpler and cleaner test of whether the CP asymmetry
in $B_s\rightarrow D_s^+ D_s^-$ vanishes,
$$a_{CP}(B_s\rightarrow D_s^+D_s^-)=0.\eqn\equitest$$

Let us define the angles $\alpha$, $\beta$, $\gamma$ and
$\beta^\prime$ in a model independent way:
$$\eqalign{
\sin2\alpha\equiv a_{CP}(B\rightarrow\pi^+\pi^-),&\ \ \
\sin2\beta\equiv a_{CP}(B\rightarrow\psi K_S),\cr
\sin2\gamma\equiv a_{CP}(B_s\rightarrow\rho K_S),&\ \ \
\sin2\beta^\prime\equiv-a_{CP}(B_s\rightarrow D_s\bar D_s).\cr
}\eqn\defangle$$
The following two assumptions are practically model independent:
\item{1.} The $b\rightarrow c\bar cs$ and $b\rightarrow u\bar ud$
processes are dominated by the $W$-mediated tree diagrams.
\item{2.} In the $B^0$ abd $B_s$ systems $\Gamma_{12}\ll M_{12}$.
(This is hardly an assumption as $\Delta M/\Gamma$ is measured to
be $\sim0.7$ ($\gg1$) for $B^0$ ($B_s$), while modes that
contribute to $\Gamma_{12}$ have branching ratios of order
$\lsim10^{-3}$ ($10^{-1}$).)

With these two assumptions, the CP asymmetries in the four modes of
eq. \defangle\ always measure the phase between the mixing amplitude
and the decay amplitude (though the value of this phase may be
different in different models):
$$\eqalign{
\alpha=&\ {1\over2}\arg\left[\left({q\over p}\right)_{B^0}
\left({\bar A\over A}\right)_{b\rightarrow u\bar ud}\right],\ \ \
\beta=\ {1\over2}\arg\left[\left({p\over q}\right)_{B^0}
\left({A\over\bar A}\right)_{b\rightarrow c\bar cs}
\left({q\over p}\right)_{K^0}\right],\cr
\gamma=&\ {1\over2}\arg\left[\left({p\over q}\right)_{B_s}
\left({A\over\bar A}\right)_{b\rightarrow u\bar ud}
\left({p\over q}\right)_{K^0}\right],\ \ \
\beta^\prime=\ {1\over2}\arg\left[\left({p\over q}\right)_{B_s}
\left({A\over\bar A}\right)_{b\rightarrow c\bar cs}\right].\cr
}\eqn\Allangles$$
(In the derivation of \Allangles\ from  \defangle, one has to take into
account that $\psi K_S$ and $\rho K_S$ are CP-odd.)
With the definition of the angles through \defangle, each of the
equalities in \Allangles\ is only defined mod$(\pi$).

Within the SM, these angles are interpreted in terms of CKM phases:
$$\eqalign{
\alpha^{\rm SM}=\arg\left(-{V_{tb}^*V_{td}\over V_{ub}^*V_{ud}}\right),&
\ \ \ \beta^{\rm SM}=
\arg\left(-{V_{cb}^*V_{cd}\over V_{tb}^*V_{td}}\right),\cr
\gamma^{\rm SM}=
\arg\left({V_{ub}^*V_{ud}V_{tb}\over V_{cs}^*V_{cd}V_{ts}}\right),&\ \ \
\beta^{\prime{\rm SM}}=
\arg\left(-{V_{cb}^*V_{cs}\over V_{tb}^*V_{ts}}\right).\cr
}\eqn\SMangles$$
Furthermore, within the Standard Model,
$$\arg\left({V_{tb}^*V_{ts}\over V_{cb}^*V_{cs}}\right)=\pi+
{\cal O}(10^{-2}),\eqn\SMunit$$
leading to
$$\alpha^{\rm SM}+\beta^{\rm SM}+\gamma^{\rm SM}\approx\pi,\ \ \
\beta^{\prime{\rm SM}}\approx0.\eqn\SMpizero$$
However, from \Allangles\ we learn that {\it model-independently},
$$\alpha+\beta+\gamma-\beta^\prime=0({\rm mod}\ \pi).\eqn\sumtwopi$$
Then, obviously, $\alpha+\beta+\gamma=\pi$ is equivalent to
$\beta^\prime=0$. The sum of the three angles that in the SM
correspond to angles of the unitarity triangle will be consistent
with $\pi$
if the CP asymmetries in $B_s$ decays into final CP eigenstates
through $b\rightarrow c\bar cs$ vanish. This is independent of the
mechanism of $B^0-\bar B^0$ mixing and of whether
$\alpha,\beta,\gamma$ are related to angles of the unitarity triangle.

Two ingredients of the Standard Model are in the basis of the
prediction that $a_{CP}(B_s\rightarrow D_s^+D_s^-)\approx0$.
First, that $B_s$ mixing is dominated by box diagrams with
intermediate top quarks. Second, that CKM unitarity
(and the smallness of $|V_{ub}V_{us}|$) implies
$V_{tb}V_{ts}^*+V_{cb}V_{cs}^*\approx0$. As argued in [\nisi],
a violation of this unitarity relation always implies
large new contributions to $B_s$ mixing, either from box diagrams
with $t^\prime$ (if violation of CKM unitarity comes from a
4th generation) or from $Z$-mediated tree diagrams (if the violation
is due to a non-sequential quark). Thus, if \equitest\ is violated,
then clearly there is a significant new contribution to $B_s$
mixing. It is possible that, in addition, CKM unitarity is violated, but
that can be tested independently [\niqu].

\endpage
\centerline{\bf Ackowledgments}
I am grateful to Guy Blaylock and Abe Seiden for a very enjoyable
collaboration on the question of CP violation in $D-\bar D$ mixing.
I thank Sandip Pakvasa for very useful discussions during the
conference and for informing me on his work in preparation with
Tom Browder. This work was supported in part by the
United States -- Israel Binational Science Foundation (BSF), by the
Israel Commission for Basic Research and by the Minerva Foundation.
\vskip 2cm
\refout

\end